\documentclass[12pt,abstract=on]{scrartcl}
\usepackage{longtable}
\usepackage{ifthen}

\usepackage[colorinlistoftodos]{todonotes}
\newcommand*{\appendixmore}{%
  \renewcommand*{\othersectionlevelsformat}[1]{%
    \ifthenelse{\equal{##1}{section}}{\appendixname~}{}%
    \csname the##1\endcsname\autodot\enskip}
  \renewcommand*{\sectionmarkformat}{%
    \appendixname~\thesection\autodot\enskip}
}
\usepackage[utf8]{inputenc}
\usepackage[T1]{fontenc}
\usepackage{lmodern}
\usepackage[english]{babel}
\usepackage{amsmath, amssymb, amsthm}
\usepackage{hyperref, fullpage}
\usepackage{subfig} 
\usepackage{graphicx}
\usepackage{hyperref}
\usepackage{natbib}
\usepackage{authblk}
\usepackage{color}
\usepackage{booktabs}
\usepackage[colorinlistoftodos]{todonotes}
\newcommand{\wbp}{weak biomarker prior }
\newcommand{\sbp}{strong biomarker prior}
\newcommand{\pub}{^{(Public)}}
\newcommand{\spo}{^{(Sponsor)}}
\newcommand{\cn}{\mbox{\texttt{C}$_n$}}
\newcommand{\stn}{\mbox{\texttt{S}$_{n,\alpha_S}$}}
\newcommand{\en}{\mbox{\texttt{E}$_n$}}

\newcommand{\sstn}{\mbox{\footnotesize \texttt{S}$_{n,\alpha_S}$}}
\newcommand{\sen}{\mbox{\footnotesize \texttt{E}$_n$}}
\newcommand{\scn}{\mbox{\footnotesize \texttt{C}$_n$}}

\newcommand{\bDelta}{\mbox{\boldmath $\Delta$}}
\newcommand{\bC}{\mbox{\boldmath $C$}}

\newcommand{\bpsi}{\mbox{\boldmath $\psi$}}
\renewcommand{\sc}{{S'}}

\newcommand{\argmax}{\mbox{argmax}}

\title{Optimizing Trial Designs for Targeted Therapies\thanks{This project has received funding from the European Union’s 7th Framework Programme for research, technological development and demonstration under the IDEAL Grant Agreement no 602552, and the InSPiRe Grant Agreement no 602144.}}
\author[1]{Thomas Ondra$^\dagger$}
\author[2]{Sebastian Jobjörnsson\thanks{Contributed equally.}}
\author[3]{Robert A.~Beckman}
\author[2,4]{Carl-Fredrik Burman}
\author[1]{Franz König}
\author[5]{Nigel Stallard}
\author[1]{Martin Posch}

\affil[1]{Center for Medical Statistics, Informatics, and Intelligent Systems, Medical University of Vienna, Vienna, Austria}
\affil[2]{Department of Mathematics, Chalmers University, Gothenburg, Sweden}
\affil[3]{Departments of Oncology and of Biostatistics, Bioinformatics, and Biomathematics, Georgetown University Medical Center, Washington, D.C, US}
\affil[4]{Statistical Innovation, AstraZeneca R\&D, Molndal, Sweden}
\affil[5]{Warwick Medical School, The University of Warwick, Coventry, UK}

\date{}

\begin{document}
\maketitle
\begin{abstract}
An important objective in the development of targeted therapies is to identify the populations where the treatment under consideration has positive benefit risk balance. We consider {pivotal clinical trials}, where the efficacy of a treatment is tested in an overall population and/or in a pre-specified subpopulation. Based on a decision theoretic framework we derive optimized trial designs by maximizing utility functions. 
Features to be optimized include the sample size and the population in which the trial is performed (the full population or the targeted subgroup only) as well as the underlying multiple test procedure. The approach accounts for prior knowledge of the efficacy of the drug in the considered populations using a two dimensional prior distribution. The considered utility functions account for the costs of the clinical trial as well as the expected benefit when demonstrating efficacy in the different subpopulations. We model utility functions from a sponsor's as well as from a public health perspective, reflecting actual civil interests. Examples of optimized trial designs obtained by numerical optimization are presented for both perspectives.
\end{abstract}

\section{Introduction}

In the development of targeted therapies the investigation of potentially predictive biomarkers is  critical. If efficacy is limited to an identifiable subgroup of patients, developing a therapy for an unselected patient population is  ethically problematic and will also require unnecessarily large sample sizes because of a diluted treatment effect. On the other hand, erroneously restricting a drug development program to a subpopulation is also unethical, as it excludes patients from an effective treatment. Furthermore, it will entail a financial loss for the sponsor because of unnecessary costs of biomarker development and screening and the lower prevalence of the   future patient population.

Several one and two stage clinical trial designs have been proposed in which the treatment effect is tested in an overall population as well as in a subgroup of biomarker positive patients  \citep{mandrekar_clinical_2009-1,chen_hypothesis_2009,freidlin_randomized_2010,mandrekar_design_2011,freidlin_randomized_2012,ziegler_personalized_2012} (see \citep{ondra_methods_2016} for a recent review).  
To account for the resulting multiple comparisons, tailored multiplicity adjustments have been developed \citep{song_method_2007,alosh_flexible_2009,bretz_graphical_2009,burman_recycling_2009,Zhao_2010,spiessens_adjusted_2010,bretz_graphical_2011,millen_chain_2011,alosh_multiplicity_2013}. Alpha allocation has also been optimized using interim trial data and/or data external to the trial, with respect to a utility function, providing an early example of the use of decision analysis \citep{chen_hypothesis_2009}.

In this paper we use a comprehensive decision theoretic approach to derive optimal trial designs for the development of targeted therapies. Especially, the framework allows us to assess when it is favourable to investigate the biomarker in a clinical trial and when it is actually more efficient to disregard the biomarker and to proceed with a classical trial design. 
This extends earlier decision theoretic methods that focused on the selection of the population for clinical trials incorporating a biomarker  \citep{beckman_integrating_2011, krisam_decision_2014, Interim_Decisions_2014, kirchner_utility_2015, krisam_optimal_2015,graf_adaptive_2015}.

Consider a setting where a single potentially predictive binary biomarker has been identified in advance, separating the full population $F$ into biomarker positive ($S$) and biomarker negative ($\sc$) patients and there is prior evidence suggesting that the treatment effect may be more pronounced (or only present) in the biomarker positive group. Let $\lambda_S$ and $\lambda_\sc$, satisfying $\lambda_S + \lambda_\sc = 1$, be the prevalences of biomarker positive and biomarker negative patients in the full population. For this situation we consider three design options for a pivotal clinical trial: (i) The {\em classical design} that does not account for the biomarker status and tests for a treatment effect in the full population only, (ii) the {\em stratified design} that also recruits patients from the full population but where the biomarker status of each patient is determined and the treatment effect is tested in the full population and the subpopulation, and, (iii) the {\em enrichment design}, where patients are screened for the biomarker status and only biomarker positive patients are included in the trial. 

The choice of trial design will in general not only be based on power arguments, but on the overall expected utility of different designs, accounting for the potential rewards and costs. Rewards can be quantified by the sales revenue, from a sponsor's view, or by a measure of the overall health benefit, from a public health view. The costs of the trial  are determined by fixed and per patient costs as well as investments in biomarker development and the determination of the biomarker status for the patients in the trial. Based on a decision theoretic framework, we first optimize each of the three trial designs by choosing optimal sample sizes (and an optimized multiple testing procedure for the stratified design). Then, the optimal design can be selected among the three optimised designs based on their expected utilities. 
The optimal design choice depends on the type of utility function used (sponsor's view or public health view), the reward and cost parameters,  the prior distribution on the effect sizes and the prevalence of the biomarker positive subgroup.

\section{Testing Problem and Considered Trial Designs}\label{TheModel}

Let $\bDelta=(\delta_S,\delta_\sc)$ denote the treatment effects for the primary efficacy endpoint in the subgroup and its complement, respectively. 
Furthermore, let $\pi(\bDelta)$ denote a prior distribution on $\bDelta$. We focus on priors that satisfy $\pi(\bDelta)=0$ for $\delta_S<\delta_\sc$. This accounts for settings where there is some evidence that the effect size in the biomarker positive treatment group may be larger than in the biomarker negative group but not the other way around.

For simplicity, we assume that the basis of marketing authorization is a single pivotal trial. We further assume that a necessary condition for regulators to approve a drug for the populations $S$ or $F$ is the demonstration of a significant treatment effect in the respective population by a suitable multiple testing procedure controlling the familywise error rate (FWER) at level $\alpha$ in the strong sense.  Consider the two null hypotheses $$H_S:\delta_S\leq 0 \mbox{ and } H_F:\delta_F\leq 0,$$ where 
$\delta_F=\lambda_S \delta_S+\lambda_\sc\delta_\sc$, and let, for some trial design $d$, $\bpsi_d=(\psi_{S,d},\psi_{F,d}$) denote a multiple testing procedure such that $\psi_{i,d}=1$ (0) if there is a {statistically} significant (no significant) treatment effect in population $i=S,F$.

We consider three types of trial designs, the classical, the stratified and the enrichment design.
Let $D=\{\cn,\stn,\en|n\geq n_{\min},\alpha_S\in [0,\alpha]\}$ denote the set of considered trial designs, where $\cn,\stn,\en$ are defined below:

\paragraph{Classical design} $\cn$ refers to a classical parallel group design with per group sample size $n$ recruiting patients from the full population and testing $H_F$ only. $H_F$ is tested by a non-stratified test $\psi_{F,\scn}$ and we set $\psi_{S,\scn}=0$. 
\paragraph{Stratified Design}

 $\stn$ refers to a stratified design, which differs from the classical design in that analysis is stratified by the biomarker status and both hypotheses $H_F$ and $H_S$ are tested with a weighted multiple testing procedure with parameter $\alpha_S$. As multiple testing procedure we apply the closed  Spiessens-Debois' test \citep{song_method_2007,spiessens_adjusted_2010}. This test combines the Spiessens-Debois' test for the rejection of the intersection hypothesis $H_S \cap H_F$ with the closed testing principle so as to obtain a test for the rejection of either $H_S$ or $H_F$ (or both). Let $p_S$ and $p_F$ denote unadjusted p-values for testing $H_S$ and $H_F$, respectively. Here we assume that $H_F$ is tested with a test stratified for the biomarker (in contrast to the classical design, where a non-stratified test is used as no biomarker information is available). For $\alpha_S,\alpha_F \geq 0$, the closed Spiessens-Debois' test then rejects $H_S$ if $p_S\leq \alpha$ and either $p_S < \alpha_S$ or $p_F < \alpha_F$. Similarly, it rejects $H_F$ if $p_F\leq \alpha$ and either $p_S < \alpha_S$ or $p_F < \alpha_F$.
To control the familywise error rate at level $\alpha$ in the strong sense, the significance levels $\alpha_S$ and $\alpha_F$ need to satisfy 
\begin{equation}
P_{H_S\cap H_F}(p_S\leq \alpha_S\vee p_F\leq \alpha_F)\leq \alpha\,.\label{eq:cond}
\end{equation}
Thus, the significance level $\alpha_F$ is determined by (\ref{eq:cond})  if $\alpha_S \leq \alpha$ is given. Note that the corresponding function $\alpha_F(\alpha_S)$ depends on the subgroup prevalence $\lambda_S$.

We assume that in the stratified design, market authorization in the population $F$ is not only determined by the treatment effect in $F$, but that regulators additionally require some evidence that there is a treatment effect in both $S$ and $S'$, so that the rejection of $H_F$ is not completely dominated by a treatment effect in a single subgroup only. Thus, we assume  that the regulators' decision rule corresponds to a hypothesis test where $H_F$ is only rejected, if, in addition, the p-values $p_S$ and $p_\sc$ of tests for efficacy in the two subgroups fall below corresponding thresholds $\tau_{S}$ and $\tau_{\sc}$. The resulting modified Spiessens-Debois' test ($\psi_{S,\sstn},\psi_{F,\sstn})$ rejects $H_S$ if $\{p_S\leq \alpha\}\wedge \{p_S\leq \alpha_S \vee p_F\leq \alpha_F\}$ and  rejects $H_F$ if $\{p_F\leq \alpha\}\wedge \{p_S\leq \alpha_S \vee p_F\leq \alpha_F\}
\wedge \{p_S\leq \tau_S \wedge p_\sc \leq \tau_{\sc}\}$. Note that this test is strictly conservative, because the consistency thresholds $\tau_{S}$ and $\tau_{\sc}$ are not considered in the level $\alpha$ condition. 

\paragraph{Enrichment Design} $\en$ refers to an enrichment design, which differs from the classical design in that only patients from the subpopulation are recruited and only $H_S$ is tested. In the enrichment design, $H_S$ is tested by a test denoted by $\psi_{S,\sen}$ and we set $\psi_{F,\sen}=0$.

\section{Utility Functions}

We define utility functions that quantify the potential rewards for each of the possible trial outcomes as well as the cost of the trial. To model the rewards, we distinguish between the sponsor and the public health view, leading to different utility functions for the two perspectives:
\begin{equation}
 U^{(v)}(d) = \sum_{i=S,\sc}   \varphi^{(v)}_{i,d}- \bC_d,  \label{eq:utility}
\end{equation}
where $v=Sponsor$ for the sponsor and $v=Public$ for the public health view, $d\in D$ denotes the trial design, $\varphi^{(v)}_{i,d}$ the reward due to the trial outcome in subgroup $i=S,\sc$ and $\bC_d$ the cost of the trial.
The cost functions $\bC_d$ of the different trial designs $d \in D$ are sums of fixed costs and costs per recruited patient in the trial. Note that the per-group sample size $n$  may vary among the three designs and below we will determine optimal sample sizes for each type of design.

For the {classical design} the cost function is given by 
$$\bC_{\scn}=c_{\text{setup}} +2nc_{\text{per-patient}},$$
where the setup costs of the trial $c_{\text{setup}}$ are fixed costs and  $c_{\text{per-patient}}$  are the marginal costs per patient. 
In the stratified design there are additional fixed costs $c_{\text{biomarker}}$ to develop the biomarker and additional per patient costs to determine the biomarker status
$c_{\text{screening}}$. Thus, the cost function of the {stratified design} is given by 
$$\bC_{\sstn}=c_{\text{setup}}+ c_{\text{biomarker}} +2 n (c_{\text{per-patient}}+c_{\text{screening}}).$$
For the {enrichment design} the fixed costs are the same as for the stratified design. However, to recruit only biomarker positive patients one has to screen (on average) $2n/\lambda_S$ patients from the full population until $2n$ biomarker positive patients are identified. Given {that} the screening and determination of the biomarker status induces costs $c_{\text{screening}}$ the cost function is given by
$$ \bC_{\sen}=c_{\text{setup}}+c_{\text{biomarker}}+2n\left( c_{\text{per-patient}} + c_{\text{screening}}/\lambda_S \right).$$

\subsection{The Sponsor's Utility Function}

For the sponsor, the utility is the Net Present Value (NPV), which is defined as the reward (sales revenue) minus the trial costs. We model the sponsor's reward  as  a function of (i) the outcome of the regulatory approval process, (ii) the price the sponsor can achieve, and (iii) the size of the population the drug is licensed for.

To model (i) and (ii) we define reward functions 
$\varphi\spo_i$ for $i=S,\sc$
that specify the reward obtained in the respective population. The reward function may depend on the observed data, the design of the pivotal trial $d$ and the prevalence of the subgroup.  We model the reward as the product of the price of the drug for the treatment of a single patient times the market size. Given an overall market size $N$, the market sizes of the two subgroups are $\lambda_S N$ and $\lambda_\sc N$, respectively. Furthermore, we assume that the payers are willing to pay more if a larger treatment effect was observed. If the drug is authorized for neither subgroup, both reward functions are set to zero. If the drug is authorized for the subgroup $S$ only, the reward for the complement $\sc$ is set to zero. If the drug is authorized for the full population, we assume that the same price is charged in both subgroups. 

We assume that (given that the respective hypothesis test rejects and the observed effect size exceeds a clinically relevant threshold)  the price increases linearly with the observed effect size. Then the reward functions for the subgroups $S$ and $S'$ are 
  \begin{eqnarray}
  \varphi_{S,d}\spo&=&\left\{ 
  \begin{array}{ll}
    \lambda_S \, N \, r_S \, \psi_{S,d} \, (\hat\delta_{S,d}-\mu_S)^+ & \mbox{ if }  {\psi}_{F,d}=0\\
    \lambda_S \, N \, r_F \, \psi_{F,d} \, (\hat\delta_{F,d}-\mu_F)^+ &\mbox{ otherwise }
  \end{array}
  \right. \label{eq:price}\\
  \varphi_{\sc,d}\spo&=& \lambda_\sc \, N \, r_F \, \psi_{F,d} \, (\hat\delta_{F,d}-\mu_F)^+ ,\nonumber
  \end{eqnarray}
where $\mu_i$ denotes a minimal clinically relevant effect size for population $i = S,F$ and $(\cdot)^+$ denotes the positive part. $\hat\delta_{S,d}$ and $\hat\delta_{F,d}$ are the estimates of $\delta_S$ and $\delta_F$ obtained from the trial data. The constants $r_i$ for $i=S,F$ are the marginal prices (the change in price if the observed effect size increases by one unit) and $N$ denotes the total market size, which for the sponsor is defined as the number of future patients within the patent life of the therapy in the unselected, full population.  Note that, given that efficacy is shown in the full population, a common treatment effect estimate $\hat \delta_F$ is used in the price function. Then the overall reward within the patent life of the therapy is given by $\varphi_{S,d}\spo+\varphi_{\sc,d}\spo$.

\subsection{Public health utility function}
With the public health utility function we model the utility of trial designs under the assumption that the drug is developed by public health authorities.
Therefore, the utility of a trial is  given by the total health benefit to the society (adjusted by the production cost of the drug) minus the cost of running the trial.
We assume that the benefit of the drug is measured on a monetary scale representing the expected, accumulated (over the whole treated population) treatment effect. Costs are assumed to be the same as under the sponsor view. The reward functions for the subgroups $S$ and $\sc$ are given by

\begin{eqnarray}
  \varphi\pub_{S,d}&=&\left\{ 
  \begin{array}{ll}
     \lambda_S \, N \, r_S \, \psi_{S,d} \, (\delta_{S}-\mu_S) & \mbox{ if }  {\psi}_{F,d}=0\\
     \lambda_S \, N \, r_F \, {\psi}_{F,d} \, (\delta_F-\mu_F) &\mbox{ otherwise }
  \end{array}
  \right. \label{eq:phprice}\\
  \varphi\pub_{\sc,d}&=&  \lambda_\sc \, N \, r_F \, {\psi}_{F,d} \, (\delta_F-\mu_F) ,\nonumber
  \end{eqnarray}

The first term in the utility function (\ref{eq:utility}) denotes the total benefits summed over the whole population, which are assumed to be proportional to the effect size (adjusted for a minimal relevant threshold), if the drug is authorized. The constants $r_i$ for $i = S,F$ are the marginal benefits (the change in benefit if the effect size increases by one unit), and $N$ denotes the size of the future (unselected) patient population. Note that the benefit depends on the actual effect sizes $\delta_i$ and not on the corresponding trial estimates $\hat\delta_{i,d}$, implying that the benefit may be negative if the effect size is low. A consequence of this model choice is that a public health authority will take into account the risk of false positive approvals when optimizing its trial design. Such considerations are absent when a sponsor is optimizing, since we have assumed that only the estimated effects enter its utility function.

\subsection{Optimizing the expected utility}

{Recall that} $\pi$ denotes a prior distribution on the effect sizes $\bDelta$.
For a given utility function $U^{(v)}$ and set of  trial designs $D$ the design optimizing the expected utility is given by 
\begin{equation}
d^* \in \argmax_{d\in D} E_{\pi}\left[U^{(v)}(d)\right],\label{eq:dstar} 
\end{equation}
where
\begin{equation} \label{utility2} E_\pi[U^{(v)}(d)] = \int\! E_{\tiny \bDelta}[U^{(v)}(d)] \,\mathrm{d}\pi(\bDelta),
\end{equation} 
Note that the expectation is first taken over the data distribution given the effect sizes $\bDelta$ and then over the prior distribution $\pi$.

\section{Numerical Examples}\label{NumericalResults}

We consider parallel group designs for the comparison of means of a continuous outcome. We assume that the responses in the control and experimental treatment arms $k=C,T$ in subgroups $j=S,\sc$ are normally distributed with mean $\theta_{k,j}$ and variance $\sigma^2$. However, utilizing the central limit theorem, the model can be modified to account for many other situations. The mean treatment effects in the two subgroups are given by $\delta_j=\theta_{T,j}-\theta_{C,j}$. 
In the classical and the enrichment design non-stratified z-tests are performed to test $H_F$ and $H_S$, respectively. In the stratified test, an elementary p-value $p_S$ is computed from a z-test for $H_S$ based on the observations in $S$ and a p-value $p_F$ from a stratified z-test for $H_F$ stratifying by biomarker status. Then the Spiessens and Debois test is performed to adjust for multiplicity.  We set $\sigma=1$.

To be able to compute the expected utilities by numerical integration and not to have to rely on simulations, we approximated the sampling distributions for both the classical and the stratified designs by normal distributions (for the enrichment design the z-test statistic is exactly normally distributed). For the classical design, each subject recruited to the trial belongs to $S$ with probability $\lambda_S$. Therefore, each observation in group $i=T,C$ is with probability $\lambda_S$ distributed as $\text{N}(\theta_{i,S}, \sigma^2)$  and with probability $\lambda_\sc$ distributed as $\text{N}(\theta_{i,\sc}, \sigma^2)$. If the biomarker is either prognostic (such that $\theta_{C,S}\not= \theta_{C,\sc}$) or predictive (such that $\delta_S\not=\delta_\sc$) the overall treatment effect estimate $\hat{\delta}_F$ for the classical design is not exactly normal, but, for sufficiently large sample sizes, approximately normal by the central limit theorem. Because the observations are drawn from a mixture distribution, the standard deviation of $\hat{\delta}_F$ increases with the absolute differences $|\theta_{i,S}-\theta_{i,\sc}|, i=T,C$. For simplicity, in the numerical examples we assume that the biomarker is predictive only but not prognostic (i.e., $\theta_{C,S}=\theta_{C,\sc}$, see Appendix \ref{UtilitySponsor}  for further details).
For the stratified design, we assume that the subgroup estimates, $\hat{\delta}_S$ and $\hat{\delta}_\sc$, are constructed as the sample means of exactly $\lambda_S n$ (resp.~$\lambda_\sc n$) observations per group from the subgroups $S$ and $\sc$. However, if patients are not selected for the trial based on biomarker status, the number of subjects from each subgroup is binomially distributed, though, for large $n$, the random sample sizes have only little impact and the approximation becomes accurate. Therefore, in the numerical investigations, we introduced a minimal sample size of $n_{\min}=50$ patients per treatment arm. For the contour plot (Figure \ref{ContourPlot}) optimization was performed by evaluating the objective function for a grid of candidate sample sizes (and $\alpha_F$ values for the stratified design). For the optimizations in the other plots, a further optimization step was applied by optimizing the objective functions with the R Version 3.2.4 procedure {\em optim} \citep{R} using grid points as starting values.  

The one-sided significance level is set to $\alpha=0.025$ and the consistency thresholds in the multiple test for the stratified design  to $\tau_S,\tau_\sc=0.3$. We consider discrete prior distributions $\pi_{\delta_{S,i},\delta_{\sc,i}}$ on a grid $(\delta_{S,i},\delta_{\sc,i}),i=1,\ldots,l$ of effect sizes and specify two priors corresponding to scenarios where there is either only weak or strong prior evidence that the biomarker is predictive. The prior distributions used in the examples are defined in Table \ref{tab:priors} and depend on an effect size parameter $\delta$. In the examples below we set $\delta=0.3$ with the exception of Subsection \ref{overview} where optimal designs for other choices of $\delta$ are explored.

\begin{table}
$$\begin{array}{rl|llll}\hline
 & \delta_S   &0&\delta&\delta&\delta\\
& \delta_{S'}&0&0&\delta/2&\delta\\\hline
\text{''\wbp''}&  &0.2&{0.2}&0.3&{0.3}\\
\text{{''\sbp''}}& &0.2&{0.6}&0.1&0.1\\\hline
\end{array}
$$
\caption{Prior distributions corresponding to scenarios where there is either only weak or strong prior evidence that the biomarker is predictive.\label{tab:priors} The constant $\delta>0$  parametrizes the effect sizes in the prior.}\label{tab:prior}
\end{table}

The reward and cost parameters in the sponsor and the public health utility function 
are specified via the following three cases: 

\begin{description}
\item[Case 1] Corresponds to a large market, where the biomarker costs are negligible, i.e. $Nr_S=Nr_{F}=10{,}000$ Million US Dollars (MUSD) per unit of efficacy and $c_{\text{screening}}=c_{\text{biomarker}}=0$.  
\item[Case 2] Corresponds to a small market, where the biomarker costs are still negligible, i.e. 
$N r_F=N r_S=1000$ MUSD per unit of efficacy.
\item[Case 3] We add biomarker and screening costs, $c_{\text{screening}}=5000$ USD per patient and $c_{\text{biomarker}}=10$ MUSD. The reward parameters $Nr_S$ and $Nr_F$ are the same as in Case 2. 
\end{description}
For all three cases we choose $c_{\text{per-patient}} = 0.05$ MUSD and $c_{\text{setup}} = 1$ MUSD. {Note that the setup costs  are assumed to be the same for the enrichment, classical and stratified design and therefore have no impact on the order of their expected utilities. However, they do have an  impact on the sign of the utility, and thus whether any trial design is superior to no trial at all.}
In the reward functions (\ref{eq:price},\ref{eq:phprice}) we set the minimal clinical relevant thresholds to $\mu_S=\mu_F=0.1$, which is a third of the effect size $\delta=0.3$ used in the prior distributions in Section \ref{sec:strong} and \ref{sec:weak}. 

\subsection{Results}\label{Results}
We discuss the optimal designs for the weak and the strong biomarker prior and the three cases specifying the cost and reward parameters. 
\subsubsection{Optimization under the Weak Biomarker Prior \label{sec:weak}}
\begin{figure}
\includegraphics[width=1 \textwidth]{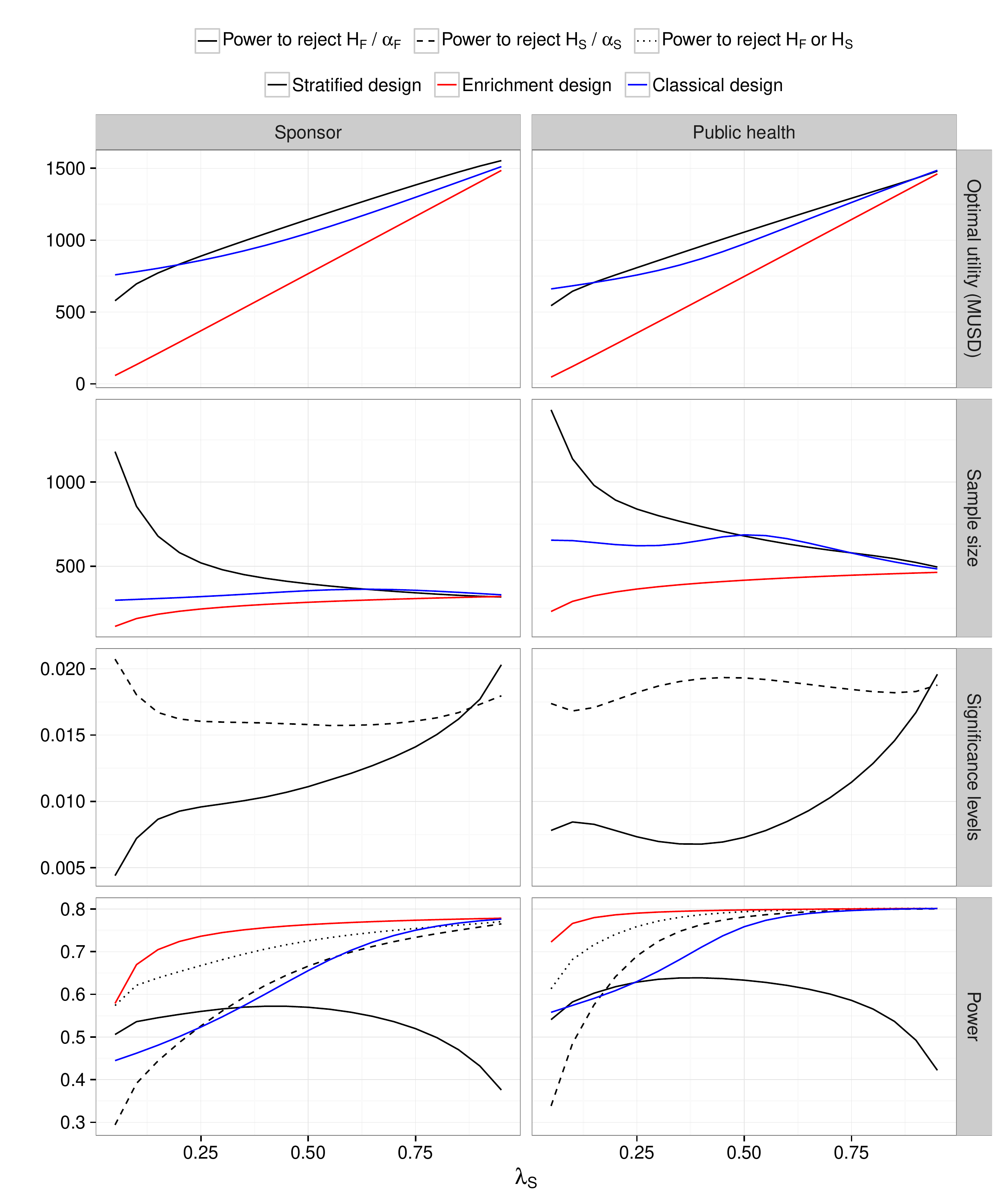}
\caption{ Weak biomarker prior and a large market with no biomarker costs (Case 1). Optimized expected utilities and sample sizes for the enrichment, classical and stratified design as functions of the prevalence for  {$\lambda_S\in [0.05,0.95]$}. For the stratified design, optimized levels $\alpha_S$ and $\alpha_F$ for the multiple testing procedure are given. The last row shows the overall probability (averaged over the prior) that a significant treatment effect in $H_S$ or $H_F$ can be shown (and, for the stratified design, that the thresholds $\tau_S$ and $\tau_\sc$ are crossed). The priors are defined as in Table \ref{tab:priors} with $\delta=0.3$.}
\label{sCc1}
\end{figure}

\paragraph{Large market, no biomarker costs (Case 1)} 
The optimized utilities and corresponding optimal classical, stratified and enriched designs are shown in Figure \ref{sCc1}. 

\begin{figure}
\includegraphics[width=1 \textwidth]{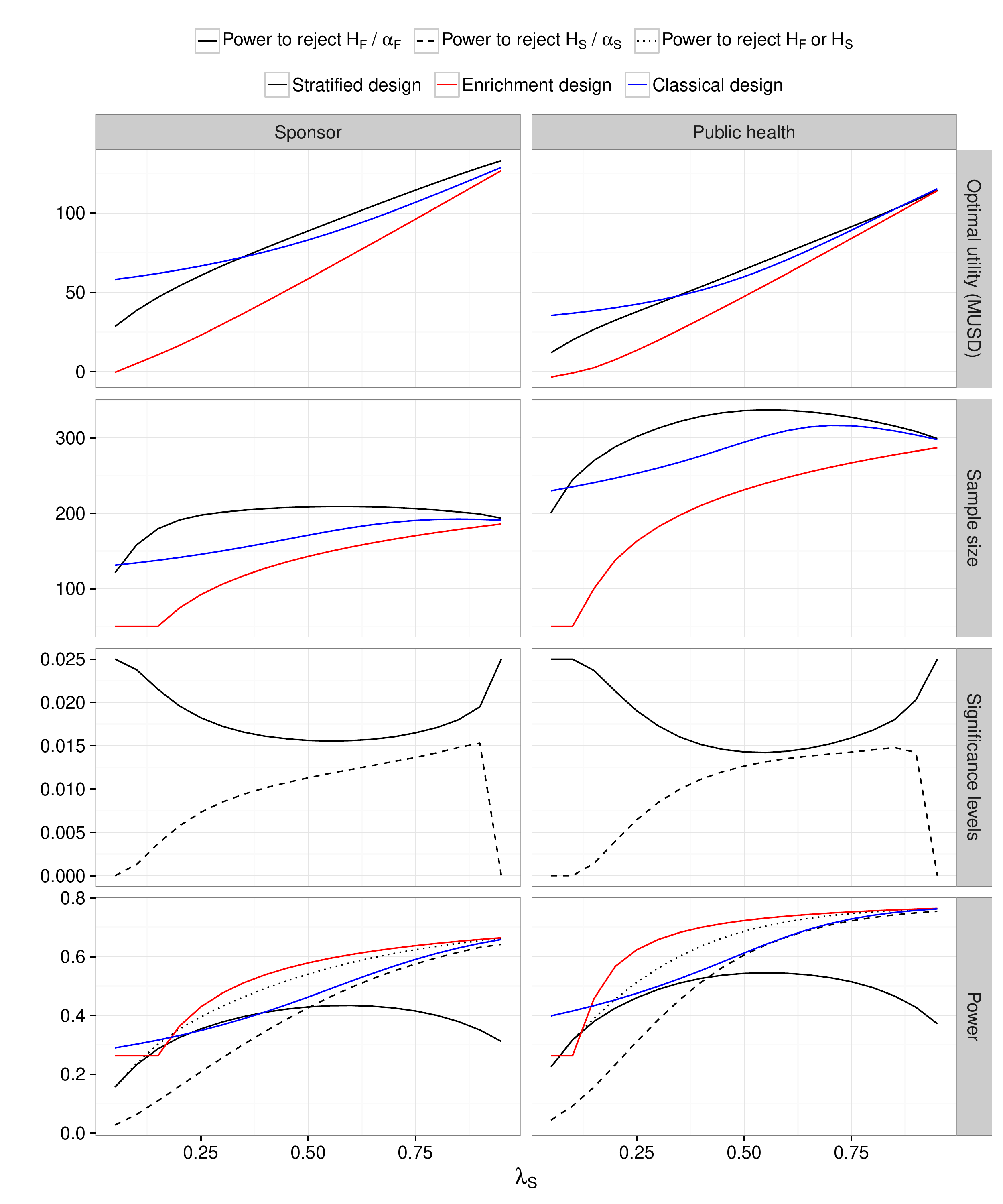}
\caption{Weak biomarker prior and a small market with no biomarker costs (Case 2). See the legend of Figure \ref{sCc1}.}
\label{sCc2}
\end{figure}

\begin{figure}
\includegraphics[width=1 \textwidth]{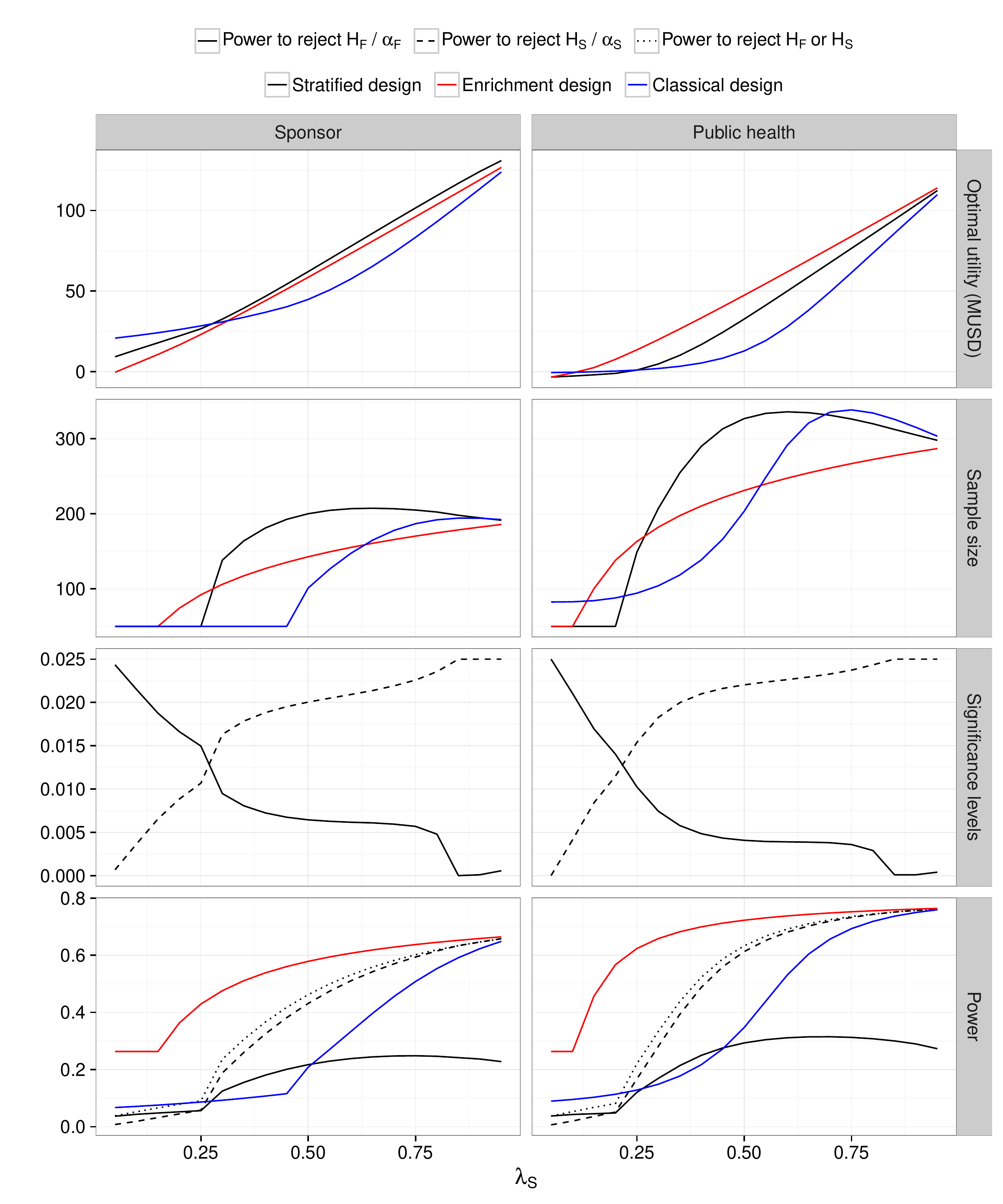}
\caption{ Strong biomarker prior and a small market with no biomarker costs (Case 2). See the legend of Figure \ref{sCc1}.}
\label{sAc2}
\end{figure}

{\em Optimal utility.} For the sponsor utility function, the stratified design has the largest expected utility, except for low prevalences where the classical design is optimal. The latter is on first sight surprising, because in Case 1 we assume no biomarker costs. However, in  the stratified design (in contrast to the classical design), to show efficacy in the full population, we require that $p_S$ and $p_{S'}$ do not exceed $\tau_S= \tau_{S'}=0.3$ (in addition to rejection of $H_F$ in the multiple testing procedure). Thus, for low prevalences the sample size of the stratified design needs to be substantially increased to reach a sufficient power to show efficacy in $F$ and therefore its expected utility is lower. 

For the public health utility function we observe a similar pattern. However, for large $\lambda_S$ the expected utility of the classical design is almost identical to that of the stratified design. This holds because the power to reject $H_S$ in the optimized stratified design approaches the power to reject $H_F$ in the classical design and the rewards obtained for authorization in populations $S$ and $F$ are similar. Why is the stratified design for the sponsor view still optimal in this case? This results from the fact that the size of the reward in the sponsor view depends on the observed rather than the true treatment effect: for trial outcomes where $H_F$ can be rejected in the classical design but, due to the variability of estimates, $\hat \delta_S$ is large but $\hat \delta_\sc$ is small (and thus $\hat \delta_F<\hat \delta_S$) the reward for a market authorization in $S$ may become larger than the reward in $F$. However, while the classical design leads to rejection of $H_F$ in such cases, the stratified design rejects $H_S$ and not $H_F$ because of the consistency threshold.

{\em Optimal sample size.} Overall, the optimized sample sizes for the public health utility function are larger than for the sponsor utility function. They are lowest for the enrichment design, and - for smaller prevalences - largest for the stratified design. For the latter, the sample size increases sharply for low prevalences. This is due to the fact that a sufficient sample size in the subgroup is required to achieve adequate power for the rejection of both $H_S$ and $H_F$ (for the latter due to the consistency threshold $\tau_S$).
Furthermore, the relationship of the optimal sample size and the prevalence is qualitatively different for the three designs. For both utility functions the optimal sample size is increasing in the prevalence for the enrichment design (because the gain when demonstrating efficacy in $S$ increases), decreasing for the stratified design (because, as noted above, a sufficient sample size in $S$ is required for the rejection of $H_S$ and for the rejection of $H_F$) and non-monotone for the classical design (essentially because the effect size in population $F$ is increasing in $\lambda_S$ such that for small $\lambda_S$ the expected utility does not sufficiently increase with the sample size to compensate the additional costs, while for large $\lambda_S$ a smaller sample size is sufficient to achieve adequate power).

{\em Significance levels.}
{In the intersection hypothesis test of the optimal multiple testing procedure in the stratified design $\alpha_S$ is larger than $\alpha_F$ for almost all prevalences. To make up for the lower sample sizes in the subgroup, the optimal design uses a larger $\alpha_S$ than $\alpha_F$. For increasing prevalences, the correlation of the test statistics used to test $H_S$ and $H_F$ increases such that less multiplicity correction is required and both $\alpha_S$ and $\alpha_F$ increase.

{\em Power.}
We define the power corresponding to a specific trial design as the overall probability (averaged over the prior) of regulatory approval in any population. This is a slight generalization of the traditional concept of power, which in the current context may be defined as the probability of regulatory approval conditional on a specific pair of subgroup effects. The power obtained by averaging over a prior has also been referred to as \emph{assurance} \citep{ohagan2005assurance}. The curves shown in Figure \ref{sCc1} correspond to the optimal designs. It can be seen that the power is largest for the enrichment design, followed by the stratified and the classical design and that it increases with the prevalence.

Note that for the stratified design, the probability to obtain marketing authorization in $H_F$ is largest for intermediate values of $\lambda_S$ and much lower than for the classical design if $\lambda_S$ is large (even though the optimized sample sizes are similar in this case). This is due to the application of the consistency thresholds which are a more difficult to meet if one of the subgroups $S$ or $\sc$ is small.

\paragraph{Small market, no biomarker costs (Case 2)} 
Case 2 differs from Case 1 only in that the rewards $Nr_F$ and $Nr_S$ are reduced by a factor $10$.
Because of the lower rewards the optimized expected utilities are smaller compared to Case 1 (see Figure \ref{sCc2} for the expected utilities and optimized design parameters). They  decrease even more than by a factor $10$ as the trial costs are not reduced proportionally.
However, the optimized sample {sizes} (and consequently the overall probabilities to show efficacy in the respective populations)  are substantially smaller than in Case 1. Overall, the expected utilities follow a similar pattern as in Case 1 but the range of prevalences where the classical design has a higher expected utility is larger than in Case 1 for both the sponsor and the public health utility functions. The assumption of a smaller market qualitatively changes the optimized sample size of the stratified designs as a function of the prevalence. For low prevalences the optimized sample size is much lower than in Case 1: because the reward is lower, it does not pay off to invest in a large overall sample size to meet the threshold $\tau_S$ in the subpopulation. This is also reflected in the optimized significance levels $\alpha_S$ and $\alpha_F$, which give more weight to $H_F$ than in Case 1. 

\paragraph{Small market with biomarker costs (Case 3)}
Note that the addition of biomarker costs has no impact on the expected utility of the optimal classical design (as it does not require the biomarker). However, the expected utilities of the enrichment and the stratified design become smaller compared to Case 2 because of the additional costs. Therefore, the classical design now dominates the stratified design for a broader range of (small) values of $\lambda_S$ and the stratified design becomes optimal only for larger values of $\lambda_S$ (see Figure \ref{sCc3} in the Supplementary Material). In the public health view, the classical design dominates the stratified design also for very large values of $\lambda_S$: even though the classical design leads to lower expected rewards compared to the stratified design (since the latter is more likely to lead to market authorization for too large a population), this is compensated by the lower costs because no biomarker is required. In the sponsor view in contrast, the difference between the expected rewards of the stratified and the classical design is larger because it is determined by observed treatment effects (see also the discussion of expected utilities in Case 1, where a similar pattern is observed). Therefore, the stratified design dominates also for large values of $\lambda_S$. 

Moreover, the biomarker costs lead to a reduction in sample size compared to Case 2.

\subsubsection{Optimization under the Strong Biomarker Prior \label{sec:strong}}
First, note that the expected utility and optimal sample size of the enrichment design is the same for the weak and the strong biomarker prior because the prior distribution on the treatment effect in $S$ is identical in both. 

\paragraph{Large market, no biomarker costs (Case 1)} For the  sponsor's utility function the stratified design is still optimal, with the exception of very low prevalences (see Figure \ref{sAc1} in the Supplementary Material}). In contrast, for the public health utility function, the enrichment and the stratified design have almost identical expected utilities unless the prevalence is small. 

\paragraph{Small market, no biomarker costs (Case 2)} While, as in Case 1, the stratified design is optimal for the sponsor view for all but very low prevalences, the difference between the expected utilities of the stratified and enrichment design is small.

In contrast, for the public health utility function the enrichment design achieves the highest expected utility (see Figure \ref{sAc2}). Furthermore, for very low prevalences, none of the trial designs has a positive expected utility in the public health view and the optimal strategy is to perform no trial at all.
For the sponsor view it is still optimal to perform a trial in the unselected population, albeit with the minimal sample size if the prevalence is small. This is due to the assumption that the NPV depends on the observed effect sizes, which implies that the sponsor benefits from a high variability of the treatment effect estimates. 

Note that the optimal test in the stratified design gives most weight on $H_F$ for low and on $H_S$ for large prevalences. This holds for both the public health and the sponsor utility function.

\paragraph{Small market, biomarker costs (Case 3)}
The pattern is very similar to Case 2, however, the range of $\lambda_S$ values where the classical (for the sponsor utility) or no trial (for the public health utility) are optimal becomes larger (see Figure \ref{sAc3} in the Supplementary Material).

\subsubsection{Optimized Designs for Varying Effect Sizes\label{overview}}

\begin{figure}
\includegraphics[width=1 \textwidth]{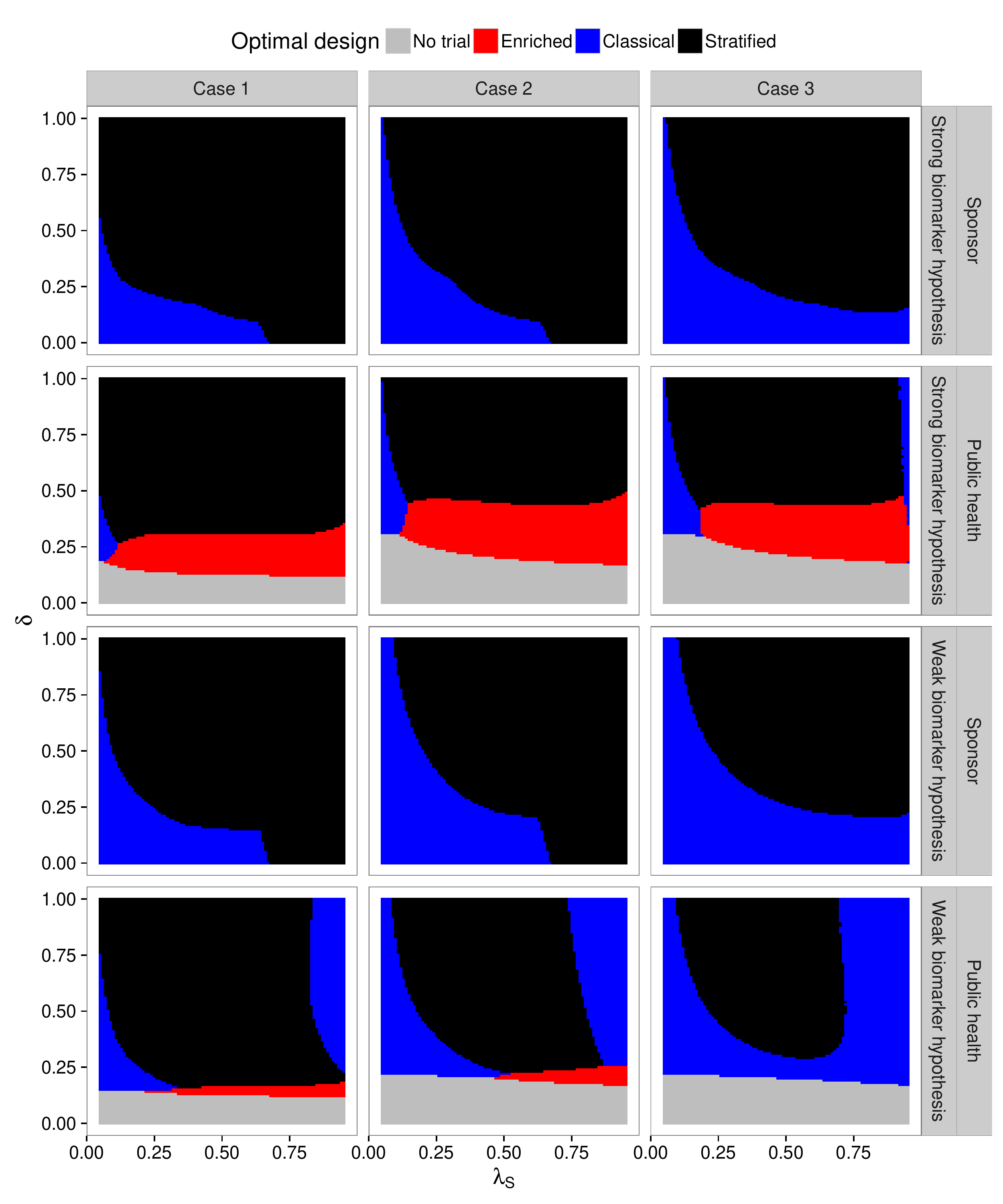}
\caption{Optimal designs for different combinations of the prevalence  $\lambda_S\in [0.05,0.95]$ and effect size parameter  $\delta\in[0,1]$. Optimized utilities for the sponsor and the public health authority are shown for both the weak and the strong biomarker prior (as defined in Table \ref{tab:priors}) under the three different cost structures defined by Cases 1, 2 and 3. The colour in a specific point indicate the type of the optimal design. Grey areas correspond to regions where all optimized designs have negative utilities, implying that the optimal choice is to perform no trial.}
\label{ContourPlot}
\end{figure}
Figure \ref{ContourPlot} shows the optimal design  as function of the prevalence $\lambda_S$ and the effect size parameter $\delta$ which parametrizes the effect sizes in the weak and strong biomarker priors in Table \ref{tab:priors}. Under the sponsor view, either the classical or the stratified design is optimal while the enrichment design never maximizes the expected utility.  Surprisingly, even for $\delta=0$, it is never optimal from a sponsor point of view to conduct no study at all in the scenarios investigated. This is due to the fact that a false positive, even though unlikely, may lead to a large reward. 
Therefore the optimal sample size is the minimal sample size $n_{\min}$ in these scenarios. This choice minimizes the costs and maximizes the variability of estimates.

For the public health view in contrast, for very low effect sizes, the optimal decision is to perform no trial at all. Under the weak biomarker prior, the enrichment design is optimal under the public health view only in the scenarios without biomarker costs, for small $\delta$ and large enough prevalences (such that the population that will benefit from a new treatment in the future is large enough). For larger effect sizes the classical design is optimal for very low and very large prevalences and the stratified design otherwise. 

Under the strong biomarker prior and intermediate $\delta$, the public health utility is optimized by the enrichment design (unless  the prevalence is too low and the classical design dominates). For larger $\delta$ the stratified design is optimal, again with the exception of very low prevalences. In addition, in the scenario with biomarker costs the classical design becomes optimal for large prevalences.

\section{Discussion\label{Discussion}}
The current study suggests decision-theoretic models for optimizing confirmatory biomarker trials, both from a sponsor and a public health perspective. Furthermore, it explores the potential discrepancies between the two perspectives. 

The optimized designs depend sensitively on the particular configuration of parameter values. Besides the priors on the effect sizes, the assumptions on the market size and costs have a substantial impact on the optimized designs. Therefore, formulating simple rules of thumb for trial designs is hardly feasible. However, a few general observations can be made. The optimized sample sizes for the public health utility function are consistently larger than for the sponsor utility (assuming the same costs, market size and reward parameters $r_F,r_S$ in both utility functions). This finding is likely due to the fact that sponsor benefit is based on the estimate of the benefit in the trial, whereas the public health benefit depends on the actual benefit. Thus the public health perspective implies a higher standard for the evidence. This finding provides a quantitative basis for the qualitative observation that health authorities tend to require a higher standard of evidence than desired by some sponsors. Mechanism design theory could potentially be applied to try to create mechanisms which align the incentives more completely.

Furthermore, for very low prevalences, the classical design outperforms the designs that are based on the biomarker. However, in these scenarios the expected utility for all designs can be negative in the public health perspective and so weakly positive in the sponsor perspective that the sponsor would allocate its resources elsewhere as well.

We find that in the sponsor view the enrichment designs never maximize the NPV in the considered scenarios. This is due to the fact that the sponsor may benefit from an authorization in the larger population even if the treatment is effective in the subpopulation only. For similar reasons, even under the global null hypothesis the strategy to perform a trial (with minimum sample size) gives a positive NPV in the sponsor view (a phenomenon that was observed also in other contexts \citep{posch2013adaptive}).

In the public health view the enrichment designs are optimal for a range of scenarios. Especially, if there is sufficient prior medical understanding that the biomarker negative  subpopulation is unlikely to be positively affected by the drug, it can be a waste of resources to conduct the trial in this population. Ethical considerations reinforce this, as it can be argued that genuine informed consent \citep{burman_carlberg_2010} implies that patients should not be randomised if their expected utility is higher on standard of care than on randomised trial medication. On the other hand, in particular when subpopulations can be expected to be similar in efficacy, it is not always worthwhile to conduct biomarker screening. In fact, there is an increased risk in a stratified trial that the treatment is rejected in the biomarker negative subpopulation due to chance. Still, in situations with genuine uncertainty about the relative efficacy in the two subpopulations, biomarker determination and stratified designs may have a large value. An obvious extension of our model is to allow for trial adaptations, potentially closing the biomarker negative part of the trial at an interim, in case results are negative \citep{brannath2009confirmatory,bauer2016twenty}.

When applying the presented framework to practical design decisions, the different model components should be scrutinized. In the numerical 
example we have assumed for simplicity that the biomarker is not prognostic but in practice this will often not be the case. If the biomarker is also prognostic, the variability of the effect size estimates will be increased with a consequent decrease in the expected utility of the classical design. 

As regards the market size for the sponsor, $N$ denotes the number of patients, determined by the patent life, for which full payment will be received upon regulatory approval. On the other hand, for the public health authority, $N$ denotes the total number of future patients. In an extended model, $N$ could be fixed to always be the total number of future patients and a factor could be added next to $N$ for the sponsor. This factor would then represent the fraction of patients corresponding to the patent life, and could be made to depend on the choice of trial design in various ways.  For example, in the enrichment design we accounted only for the screening costs arising from the determination of the biomarker status of patients. However, if the restriction of the trial population leads to slower recruitment and consequently a later authorization of the drug, the result will be a reduction of the market size and the remaining patent life. This, in turn, may reduce the potential reward in different ways for the two perspectives. Another simplification made in our framework is the assumption of a zero discount rate for the sponsor. In practice, a commercial sponsor would use a non-zero rate to discount future revenues, which would lead to a further reduction of its expected utility as compared to a public health authority.

In the considered model we assumed that the subgroup prevalences in the trial are the same as in the total patient population. However, unless the recruitment is stratified by subgroup, the actual prevalence in the trial will be stochastic.  Furthermore, the propensity to participate in the trial may vary between subpopulations. While our results are generally robust to random variations in the prevalence, varying propensities for trial participation may lead to a biased estimation of the effect size in the full population in the classical design (and also the stratified design, if an overall effect estimate based on observed prevalences is computed). The question of generalizability of trial results to general patient populations is however not specific to the development of targeted therapies but a more general issue.

We did not explicitly incorporate a benefit risk evaluation of the treatment into the model. However, the parameters $\mu_S$ and $\mu_F$ in the reward functions (\ref{eq:price},\ref{eq:phprice}) can be interpreted as the minimal treatment effects that compensate the ''costs'' of the treatment, as the burden of treatment, side effects and monetary costs. While these are considered as given in our model, they could alternatively be estimated from clinical trial data. 

We modelled the sponsor and public health utility as essentially linear functions of the observed and true effect size, respectively.
From a commercial perspective this can be reasonable for scenarios where no alternative treatment options are available.  However, if competitor products are on the market, the model may need to be modified because the market share, in terms of number of doses prescribed, and not only the price or benefit per patient may depend on the effect size. This can be incorporated by models where the market share is a function of the posterior distribution of efficacy (and possibly safety) parameters \citep{gittins_pezeshk_2000, kikuchi_pezeshk_gittins_2008}. Another aspect of our model for reimbursement concerns the pricing. Although NICE in the UK indicates that they, in our situation, would accept a price proportional to net benefit, payers in other countries may use other price models, possibly closer to a constant price. As an alternative to our linear sales model, an aggregated commercial model could be plugged in and similar optimization could be performed.

Finally, we note that the case of a very low prevalence, small market size and no biomarker costs mimics the situation of a rare disease, except there is no complementary subgroup $\sc$. Therefore, our results could be seen to suggest that the investigation of rare diseases is not recommended in either
perspective. Consequently, the question arises if research in rare diseases should receive special priority and be subsidised by society such that drug development occurs even though the expected utility to society is negative, or  in some case weakly positive but less positive than other alternative expenditures. However, this argument raises ethical questions because the
purely utilitarian viewpoint that underlies the decision theoretic framework does not account for other ethical principles as fairness and justice. Similar issues arise for small subgroups of common diseases, an increasing issue in cancer given the fact it being subdivided into many small molecular subclasses. In the case
of cancer, increased benefit due to matching between molecular subgroups and targeted therapies may mollify this issue, but this remains to be seen in individual cases, so that the ethical and public policy dilemma may still be present.

\bibliography{InSPiRe}{}
\bibliographystyle{chicago}

\newpage

\appendix

\section*{Appendix}
\section{Computation of Expected Utilities} \label{UtilitySponsor}
We derive the expected utilities for a given effect size $\bDelta$ for the enrichment, the classical and the stratified design. The overall expected utilities are then obtained from \eqref{utility2} by integrating over the prior distribution.

\paragraph{Enrichment Design} For the enrichment design, $\psi_{F,\sen}=\varphi^{(Sponsor)}_{\sc,\sen}=0$. Thus, the utility is given by
$$ U\spo(\en) = \varphi_{S,\sen}\spo - \bC_{\sen}  =\lambda_S \, N \, r_S \, \psi_{S,\sen} \, (\hat\delta_{S,\sen}-\mu_S)^+-\bC_{\sen}.$$ 
Integrating over the resulting truncated normal distribution, the expected utility given $\bDelta$ is given by
\begin{equation}\label{UtilityEnrichment}
E_{ \tiny \bDelta }[U^{(Sponsor)}(\en)] = \lambda_S \, N \, r_S \left( (1-\Phi(\kappa)) (\delta_{S} - \mu_S) + V[\hat\delta_{S,\sen}]^{1/2} \phi(\kappa) \right) -\bC_{\sen}, \notag
\end{equation} 
where $\phi$ denotes the density and $\Phi$ denotes the cumulative distribution function of the standard normal distribution, $V[\hat\delta_{S,\sen}]=2\sigma^2/n$ and
\begin{equation}
\kappa=\left[\max\left(
 \Phi^{-1}(1-\alpha) V[\hat\delta_{S,\sen}]^{1/2},\mu_S\right)-\delta_{S}\right]/V[\hat\delta_{S,\sen}]^{1/2} . \notag
\end{equation}
Similarly, for the public health view utility function we obtain
$$ E_{ \tiny \bDelta }[U^{(Public)}(\en)] =\lambda_S \, N \, r_S \, (\delta_S - \mu_S)\left( 1 - \Phi \left( \Phi^{-1} \left(1-\alpha \right) - \delta_S \, V[\hat\delta_{S,\sen}]^{-1/2} \right)  \right) - \bC_{\sen}.$$  

\paragraph{Classical Design} In the classical design, $\psi_{S,\scn}=0$, $\varphi^{(Sponsor)}_{S,\scn}=\varphi^{(Sponsor)}_{\sc,\scn}$ and 
$$U^{(Sponsor)}(\cn)= \varphi^{(Sponsor)}_{S,\scn} +  \varphi^{(Sponsor)}_{\sc,\scn} - \bC_{\scn} = N \, r_F \, \psi_{F,\scn} \, (\hat\delta_{F,\scn}-\mu_F)^+-\bC_{\scn} $$

If the mean response in $S$ and $\sc$ differ, it follows that the observations in the experimental treatment and control group follow a mixture distribution of two normal distributions. Therefore, the variance of $\hat\delta_{F,\scn}$ in the classical design is given by $$V[\hat\delta_{F,\scn}]=(2\sigma^2+\lambda_S \, \lambda_\sc \, ((\theta_{T,S}-\theta_{T,\sc})^2+(\theta_{C,S}-\theta_{C,\sc})^2))/n.$$
Thus, the expected utility given effect sizes $\bDelta$ for the classical design is given by 
$$E_{ \tiny \bDelta }[U^{(Sponsor)}(\cn)] = N \, r_F \, \left( (1-\Phi(\kappa)) (\delta_F - \mu_F) + V[\hat \delta_{F,\scn}]^{1/2} \phi(\kappa) \right) - \bC_{\scn},$$
where $\delta_F=\lambda_S\delta_S+\lambda_\sc\delta_\sc$ and $\kappa=\left(\max( \Phi^{-1}(1-\alpha)V[\hat\delta_{F,\scn}]^{1/2},\mu_F )-\delta_F \right)/V[\hat \delta_{F,\scn}]^{1/2}$. Similarly, for the public health utility function,
$$E_{ \tiny \bDelta }[U^{(Public)}(\cn)] = N \, r_F \, (\delta_F-\mu_F)\left( 1-  \Phi \left( \Phi^{-1} \left( 1-\alpha \right) - \delta_F V[\hat\delta_{F,\scn}]^{-1/2} \right)  \right) - \bC_{\scn}. $$ 

\paragraph{Stratified Design} The utility of the stratified design is given by
$$
 U\spo(\stn)   = - \bC_{\sstn} + \begin{cases}  \lambda_S \, N \, r_S \, \psi_{S,\sstn} \, (\hat{\delta}_{S,\sstn} - \mu_S)^+ & \text{if } {\psi}_{F,\sstn}=0 \\ N \, r_F \, {\psi}_{F,\sstn} \, (\hat{\delta}_{F,\sstn} - \mu_F)^+  & \text{otherwise.}\end{cases} 
 $$
The utility of the stratified design depends on the stratified treatment effect estimate in the full population (in the following we shorten the notation by dropping the design index, $\hat{\delta}_{F}:=\hat{\delta}_{F,\sstn},\hat{\delta}_{S}:=\hat{\delta}_{S,\sstn},\hat{\delta}_{\sc}:=\hat{\delta}_{\sc,\sstn},\psi_F:=\psi_{F,\sstn},\psi_S:=\psi_{S,\sstn}$) which is a weighted sum of $\hat{\delta}_{S}$ and $\hat{\delta}_{\sc}$. The expected utility given the effect sizes $\bDelta$ is given by
\begin{equation*}
E_{ \tiny \bDelta }[U\spo(\stn)] = N \, r_F \, E_{ \tiny \bDelta }\left[ {\psi}_{F} \left(\hat{\delta}_{F} - \mu_F \right)^+  \right] + \lambda_S \, N \, r_S \, E_{ \tiny \bDelta } \left[ \left( 1 - {\psi}_{F} \right) \psi_{S} \left(\hat{\delta}_{S} - \mu_S \right)^+ \right] - \bC_{\sstn}.
\end{equation*}
and can be computed by numeric integration: Let $A_F \left( n, \alpha_S ; \sigma, \alpha, \lambda_S, \tau_S, \tau_{S'}, \mu_F \right)$ denote the region in $\mathbb{R}^2$ where $\psi_{F} (\hat{\delta}_{F} (Z_S, Z_{S'}) - \mu_F )^+ > 0$ and let $A_S \left( n, \alpha_S ; \sigma, \alpha, \lambda_S, \tau_S, \tau_{S'}, \mu_S \right)$ be the region where $( 1 - \psi_{F} ) \psi_{S} (\hat{\delta}_{S} (Z_S) - \mu_S)^+ > 0$, where  $\hat{\delta}_{F} (Z_S, Z_{\sc})$ is the stratified treatment effect estimate and $Z_S,Z_\sc$ the z-statistics computed from the observations in $S$ and $\sc$ respectively. Then
\begin{equation*}
E_{  \mbox{\tiny$\bDelta$}}\left[ {\psi}_{F} \left(\hat{\delta}_{F}(z_S, z_{S'}) - \mu_F \right)^+  \right]  =  \iint\limits_{A_F} \! \left( \hat{\delta}_{F} (z_S, z_{S'}) - \mu_F \right) \phi(z_S) \phi(z_{S'}) \, \mathrm{d} z_S \, \mathrm{d} z_{S'},
\end{equation*}
and 
\begin{equation*}
E_{  \mbox{\tiny$\bDelta$} } \left[ \left( 1 - {\psi}_{F} \right) \psi_{S} \left(\hat{\delta}_{S} (Z_S) - \mu_S \right)^+ \right]  = \iint\limits_{A_S} \! \left( \hat{\delta}_{S} (z_S) - \mu_S \right) \phi(z_S) \phi(z_{S'}) \, \mathrm{d} z_S \, \mathrm{d} z_{S'} .
\end{equation*}
The shapes of the regions $A_F$ and $A_S$ depend on the specific values of the parameters and the design variables ($\alpha_S,\alpha_F,\tau_S, \tau_\sc$ and $n$). However, the regions may in all cases be described by means of a finite number of straight lines, implying that the expected values above can be computed using standard software for numerical quadrature in two dimensions. But since the integrands are linear in $z_{S'}$ and $Z_{S'}$ follows a normal distribution, one-dimensional integration may be carried out analytically in the $z_{S'}$-direction before applying a numerical method. This leads to faster numerical evaluations, which is useful when investigating how the optimal solution changes over the parameter space.

For the public health view the expected utility given $\bDelta$ may be written as
\begin{equation*}
E_{ \tiny \bDelta }[U^{(Public)}(\stn)] = N \, r_F \, (\delta_F-\mu_F) \, E_{ \tiny \bDelta } \left[ {\psi}_{F} \right] +  \lambda_S \, N \, r_S \, (\delta_S-\mu_S) \, E_{ \tiny \bDelta } \left[ \left( 1 - {\psi}_{F} \right) \psi_{S}\right] - \bC_{\stn}.
\end{equation*}
The numerical evaluation is similar to the evaluation of the conditional expectation of the utility of the stratified design for the sponsor's view. 

\newpage
\setcounter{equation}{0}
\setcounter{figure}{0}
\setcounter{table}{0}
\setcounter{page}{1}
\makeatletter
\renewcommand{\theequation}{S\arabic{equation}}
\renewcommand{\thefigure}{S\arabic{figure}}
\renewcommand{\bibnumfmt}[1]{[S#1]}
\renewcommand{\citenumfont}[1]{S#1}

\begin{center}
\textbf{ \Large Supplementary Material for \\
Optimizing Trial Designs for Targeted Therapies}
\large Thomas Ondra, Sebastian Jobjörnsson, Robert A. Beckman,
Carl-Fredrik Burman, Franz König, Nigel Stallard, and
Martin Posch
\end{center}
\begin{figure}
\includegraphics[width=1 \textwidth]{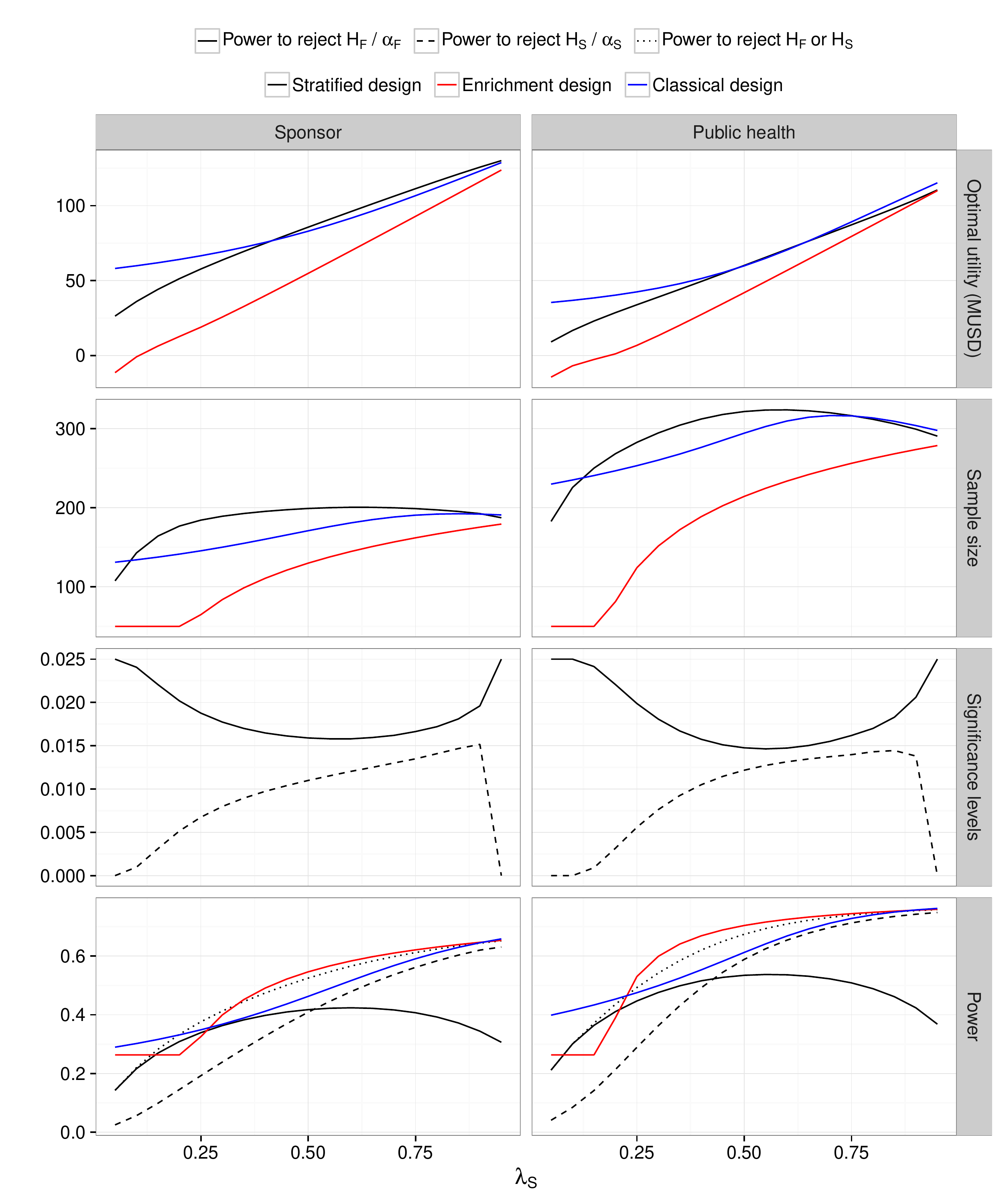}
\caption{ Weak biomarker prior and a small market with biomarker costs (Case 3). Optimized expected utilities and sample sizes for the enrichment, classical and stratified design as functions of the prevalence for  $\lambda_S\in [0.05,0.95]$. For the stratified design, optimized levels $\alpha_S$ and $\alpha_F$ for the multiple testing procedure are given. The last row shows the overall probability (averaged over the prior) that a significant treatment effect in $H_S$ or $H_F$ can be shown (and, for the stratified design, that the thresholds $\tau_S$ and $\tau_\sc$ are crossed).}
\label{sCc3}
\end{figure}
\begin{figure}
\includegraphics[width=1 \textwidth]{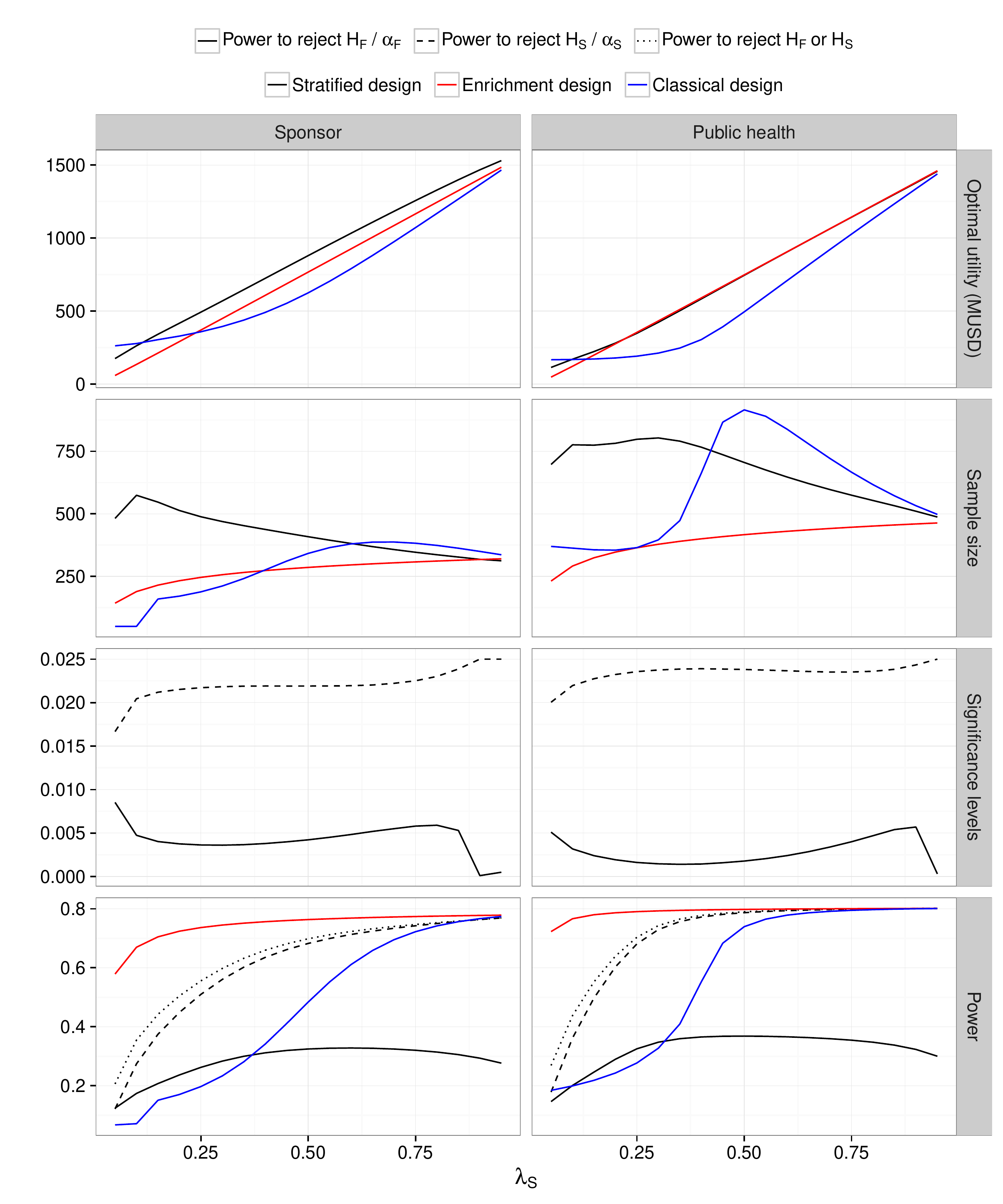}
\caption{ Strong biomarker prior and a large market with no biomarker costs (Case 1). See the legend of Figure \ref{sCc3}.}
\label{sAc1}
\end{figure}
\begin{figure}
\includegraphics[width=1 \textwidth]{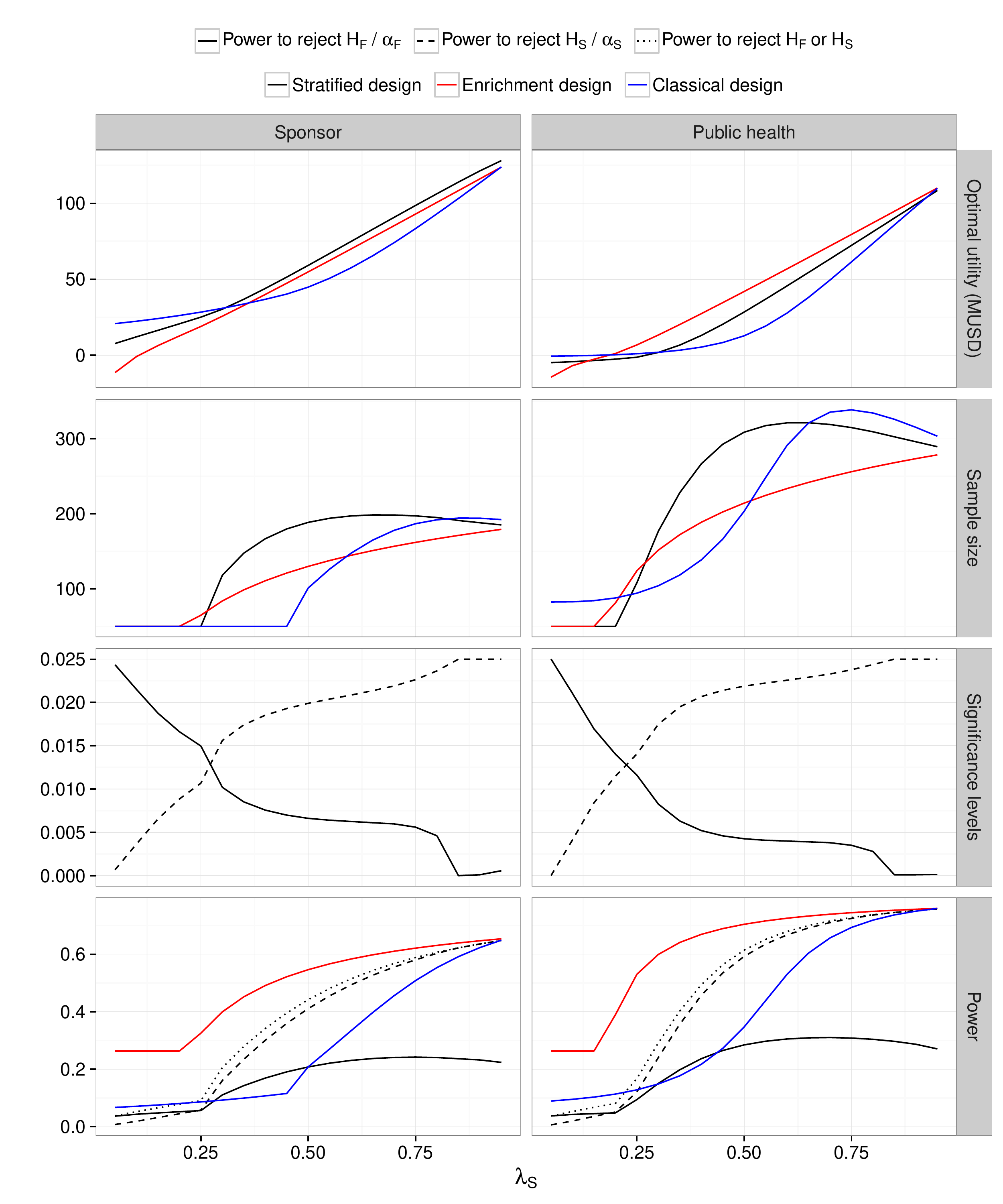}
\caption{ Strong biomarker prior and a small market with biomarker costs (Case 3). See the legend of Figure \ref{sCc3}. }
\label{sAc3}
\end{figure}

\end{document}